\documentclass[%
 reprint,
 amsmath,amssymb,
 aps,
 pra,
floatfix,
]{revtex4-1}

\usepackage{graphicx}
\usepackage{dcolumn}
\usepackage{amsmath}
\usepackage{bm}
\usepackage{color} 
\usepackage{braket}
\usepackage[normalem]{ulem}
\usepackage{gensymb}
\usepackage[dvipsnames]{xcolor}
\usepackage{lpic}
\usepackage{MnSymbol}
\usepackage[shortcuts]{extdash}

\DeclareRobustCommand{\orderof}{\ensuremath{\mathcal{O}}}

\newcommand{\var}[1]{\braket{(\Delta #1)^2}}

\begin{document}

\title{
Dispersive detection of radio-frequency dressed states}

\author{Sindhu Jammi}
\author{Tadas Pyragius}
\author{Mark G. Bason}
\author{Hans Marin Florez}
\author{Thomas Fernholz}
\email{corresponding author: thomas.fernholz@nottingham.ac.uk}
\affiliation{
School of Physics \& Astronomy, University of Nottingham, University Park, Nottingham NG7 2RD, UK
}

\date{\today}

\begin{abstract}
We introduce a method to dispersively detect alkali atoms in radio-frequency dressed states.
In particular, we use dressed detection to measure populations and population differences of atoms prepared in their clock states. Linear birefringence of the atomic medium enables atom number detection via polarization homodyning, a form of common path interferometry. In order to achieve low technical noise levels, we perform optical sideband detection after adiabatic transformation of bare states into dressed states. The balanced homodyne signal then oscillates independently of field fluctuations at twice the dressing frequency, thus allowing for robust, phase-locked detection that circumvents low-frequency noise. Using probe pulses of two optical frequencies, we can detect both clock states simultaneously and obtain population difference as well as the total atom number. The scheme also allows for difference measurements by direct subtraction of the homodyne signals at the balanced detector, which should technically enable quantum noise limited measurements with prospects for the preparation of spin squeezed states. The method extends to other Zeeman sublevels and can be employed in a range of atomic clock schemes, atom interferometers, and other experiments using dressed atoms.
\end{abstract}

\pacs{Valid PACS appear here}
\maketitle

\section{\label{sec:intro}Introduction}

Radio-frequency (RF) dressing of atoms in magnetic traps provides robust and very versatile control of the external degrees of freedom. This technique is used in a variety of cold-atom experiments, see~\cite{Garraway2016} for a recent review.
The dependence of the trapping potential on magnetic field amplitudes in the RF regime renders dressed traps robust against some low-frequency, environmental field noise. The first atom-chip based beam splitter for matter waves was demonstrated with this method~\cite{Schumm2005}. Versatility comes from the dependence of the trapping potential on the polarization of the RF field relative to the local static field; this provides greater design freedom compared to quasi-static magnetic traps. Experiments and proposals for interesting trap geometries include lattices~\cite{Courteille2006,Sinuco-Leon2015}, rings~\cite{Fernholz2007, Lesanovsky2007, Sherlock2011, Navez2016}, and hollow traps shaped as spheres~\cite{Colombe2004}, cylinders~\cite{Hofferberth2006}, and tori~\cite{Fernholz2007}. Species- and state-dependent control becomes possible in some scenarios~\cite{Navez2016, Bentine2017}, because the trap defining RF polarization component depends on the atomic $g$-factor. Such control provides prospects for quantum simulations of many-body physics as well atom interferometers without any free propagation~\cite{Stevenson2015}. 

In this paper, we present a method for dispersive detection of atoms that benefits directly from the intrinsic modulation of the atomic signal via phase-locked spin precession. Dispersive light-matter interaction at a very low technical noise level resulting from operation at radio-frequencies is a prerequisite for quantum-non-demolition (QND) measurements in a range of vapour cell experiments with very large atom numbers ($n\approx10^{12}$) and consequently low relative quantum noise, including spin-squeezing~\cite{Fernholz2008}, deterministic quantum memory~\cite{Jensen2011}, and teleportation~\cite{Krauter2013}. Such QND measurements also play a role in atom interferometry where it is desirable to lower the quantum projection noise~\cite{Itano1993} inherent to any atomic magnetometer~\cite{Budker2007}, clock~\cite{Ludlow2015}, or interferometer~\cite{Cronin2009} by using spin-squeezed states~\cite{Caves1981, Wineland1994} or other non-classical states~\cite{Pezze2016} as inputs.
Destructive detection methods, e.g., based on fluorescence imaging, are routinely used to achieve atomic shot noise limited detection for small~\cite{Heine2010} to large ensembles~\cite{Biedermann2009}. They are, however, not capable of generating spin squeezing needed to lower the projection noise. In contrast, dispersive measurements based upon off-resonant atom-light interactions enabled experimental demonstration of 18-20~dB spin-squeezing~\cite{Cox2016, Hosten2016}. These experiments used high-finesse optical cavities to achieve strong atom-light interaction with low atom numbers, and require significant technical effort to stabilize to sufficient robustness. In particular, measurements on standard atomic clock states, i.e. magnetic field insensitive states with magnetic quantum number $m=0$, do not seem compatible with the relatively simple low-noise techniques used with vapour cells that are based on the Faraday effect and polarimetric common path interferometry with RF sideband detection. Here, we demonstrate that another type of birefringence, the Voigt effect~\cite{Franke-Arnold2001}, can in principle be used to detect these states by similar means. We perform two-state detection to observe Rabi cycles with low technical noise and discuss prospects for achieving quantum limited performance.

More generally, the method presented here gives state-selective detection and provides additional experimental capabilities. E.g., the signal depends on the position of atoms through the resonance condition for the dressing frequency. In a system with multiple RF fields \cite{Courteille2006,Morgan2014,Harte2018} the signal gives information about the spatial distribution of the atoms \cite{Foot2017}. Methods based on either the Voigt or Faraday effect can be readily implemented in dressed atom experiments to provide low-noise detection with little additional overhead.

The paper is organized as follows. In section~\ref{sec:theory}, we describe dispersive interaction in the context of radio-frequency dressing. this treatment predicts detected signals at harmonics of the dressing frequency.
Section~\ref{sec:Experiment} reports experimental results using our method and discusses the observed noise behaviour as well as future extensions. Section~\ref{sec:conclusions} presents our conclusions. Details on dispersive interaction and quantum mechanical interaction strengths are given in appendices A and B.

\section{\label{sec:theory}Radio-frequency dressed, dispersive light-matter interaction}

\subsection{Circular and linear birefringence}

\mathchardef\mhyphen="2D 
\begin{figure}[b]
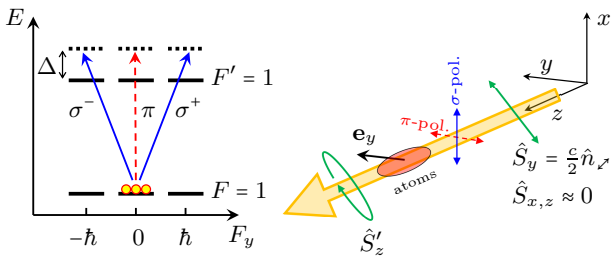

    \centering
    \begin{lpic}[]{Fig1_linear_birefringence(8.5cm)}
        \lbl[bl]{2,68,0;$E$}
        \lbl[bl]{11,54,0;$\Delta$}
        \lbl[bl]{20,42,0;$\sigma^{-}$}
        \lbl[bl]{40.5,42,0;$\pi$}
        \lbl[bl]{50,42,0;$\sigma^{+}$}
        \lbl[bl]{60,51,0;$F'=1$}
        \lbl[bl]{60,19,0;$F=1$}
        \lbl[bl]{65,5,0;$F_{y}$}
        \lbl[bl]{20,6.0,0;$-\hbar$}
        \lbl[bl]{38,6,0;$0$}
        \lbl[bl]{51,6,0;$\hbar$}
        \lbl[bl]{131,45,90;\textcolor{blue}{${\scriptstyle\sigma\mhyphen\mathrm{pol.}}$}}
        \lbl[bl]{113,39,-8;\textcolor{red}{${\scriptstyle\pi\mhyphen\mathrm{pol.}}$}}
        \lbl[bl]{100,34,0;$\mathbf{e}_{y}$}
        \lbl[bl]{112,20,24;{\tiny atoms}}
        \lbl[bl]{145,26,0;$\hat{S}_{y}=\frac{c}{2}\hat{n}_{\neswarrow}$}
        \lbl[bl]{145,16,0;$\hat{S}_{x,z}\approx0$}
        \lbl[bl]{102,3,0;$\hat{S}_z'$}
        \lbl[bl]{169,68,0;$x$}
        \lbl[bl]{153,54,0;$y$}
        \lbl[bl]{156,41,0;$z$}
    \end{lpic}
    \caption{Linear birefringence. An example level scheme for an $F=1\to F'=1$ transition is shown on the left. Only $\sigma$-transitions are allowed for atoms in $\ket{1,0}$, and corresponding polarization components of near-resonant light will acquire a phase shift proportional to atom number. An initial $45\degree$-polarization as shown on the right will become elliptical, measured by Stokes operator $\hat{S}_{z}'$ at the output.}
    \label{fig:f1det}
\end{figure}

In this section, we review the dispersive atom-light interaction arising from off-resonant laser light propagating through an atomic medium. In particular, we consider linear birefringence of an ensemble that has been prepared in a certain Zeeman sublevel, e.g., in an atomic clock state.

The basic principle can be understood by considering the simplifed example in Fig.~\ref{fig:f1det} for an atom with total spin $F=1$ and an optical transition to an excited state with $F'=1$. Off-resonant light fields experience little absorption but acquire a phase-shift proportional to transition strength and atom number. If atoms are prepared in a single Zeeman sublevel, the interactions with $\pi$- and $\sigma$-polarized fields will differ, described by different Clebsch-Gordan coefficients. For the depicted case of the quantization axis chosen along $\mathbf{e}_y$  and atoms prepared in state $\ket{F=1,F_y=0}$, the interaction with $\pi$-polarized light, i.e., linearly polarized along the $y$-axis, completely vanishes because of selection rules considering only coupling to excited states with $F'=1$. Any orthogonal polarization, however, experiences a phase shift. Light propagating along $\mathbf{e}_z$, polarized at $45\degree$ with respect to the $x,y$-axes becomes elliptically polarized, and this provides a means to measure atom number.

For a more comprehensive description of the interaction, the atomic multi-level character and arbitrary light polarization must be included. 
As detailed in Appendix~\ref{sec:dispersiveinteraction}, these can be captured by a frequency dependent polarizability tensor ${\bm{\alpha}}$ that describes the medium, and Stokes operators that describe the photon flux. 

\begin{figure}[t]
\centering
\includegraphics[width=0.48\textwidth]{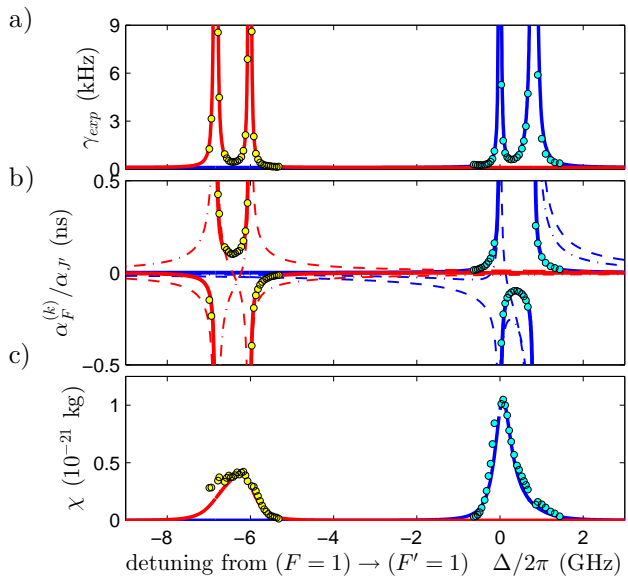}
\caption{Frequency dependence of spontaneous emission and atomic polarizability. a) Experimental decay rates $\gamma_{\mathrm{exp}}$ (circles) for both clock states using light polarized at $45\degree$ with respect to the quantization axis.
The expected behaviour (solid lines) for atoms in $F=1$ (blue group, right) and $F=2$ (red group, left) is based on measured light powers and beam sizes with 30\% correction to one of the probe lasers, possibly due to slight misalignment. b) The k-rank tensor contributions to the off-resonant D1 line polarizability (dashed, dash-dotted, solid lines for $k=0,1,2$). A single fit parameter was used to scale the measured linear birefringence (circles) to match the expected behaviour of the $k=2$-terms. c) Theoretical, off-resonant approximation and experimental data for the figure of merit $\chi=(\alpha^{(2)}_{F}/\alpha_{J'})^2\cdot I_0/\gamma$, i.e., the ratio of squared polarizability to decay coefficient (per beam intensity $I_0=P/A$), which determines the maximally achievable, signal-to-noise-power-ratio for fixed on-resonant optical density, see Eq.~\ref{eqn:explicitkappa}.}
\label{fig:alpha_pol}
\end{figure}

The dispersive interaction can be decomposed into spin dependent, irreducible tensor components of different rank $k=0,1,2$. The components are associated with corresponding polarizability contributions $\alpha^{(k)}_F$, which depend on the total spin quantum number $F$ of the atomic ground state hyperfine level.
The $^{87}$Rb atoms used in this work have nuclear spin $I=3/2$, and consequently, ground state levels with $F=1,2$. For atoms driven near the D1 lines ($J=J'=1/2$) the contributions, see general expressions in Eqs.~\ref{eq:aterms} and \ref{eq:Acontributions}, are explicitly given by 
\begin{align}
 \nonumber\alpha_{1}^{(0)}=&\frac{\alpha_{J'}}{6}\left[\frac{1}{\Delta_{1,1}} +\frac{5}{\Delta_{1,2}}\right],
 \quad\alpha_{2}^{(0)}=\frac{\alpha_{J'}}{2}\left[\frac{1}{\Delta_{2,1}} +\frac{1}{\Delta_{2,2}}\right],\\
 \nonumber\alpha_{1}^{(1)}=&\frac{\alpha_{J'}}{8} \left[\frac{-1}{\Delta_{1,1}} + \frac{5}{\Delta_{1,2}}\right],
  \quad\alpha_{2}^{(1)}=\frac{\alpha_{J'}}{8} \left[\frac{-3}{\Delta_{2,1}} + \frac{-1}{\Delta_{2,2}}\right],\\
 \alpha_{F}^{(2)}=&\frac{\alpha_{J'}}{8}(-1)^F\left[ \frac{1}{\Delta_{F,1}}+\frac{-1}{\Delta_{F,2}}\right],
\end{align}
with the far-detuned, scalar polarizability coefficient $\alpha_{J'}=\epsilon_0  \lambda_{J'}^3 \Gamma_{J'}/8\pi^2$, which depends on the D1-line parameters $\Gamma_{J'}=2\pi\times 5.75~\mathrm{MHz}$ and $\lambda_{J'}=795~\mathrm{nm}$. 
We defined detunings $\Delta_{F,F'}=\omega_L-\omega_{F,F'}$ of the light field with respect to the optical $F\rightarrow F'$ transition frequencies.
The ground- and excited state hyperfine splittings are $\Delta_{2,F'}-\Delta_{1,F'}\approx2\pi\times6835~\mathrm{MHz}$ and $\Delta_{F,2}-\Delta_{F,1}\approx2\pi\times817~\mathrm{MHz}$. This large difference relative to the small probe detuning used in our experiments, justifies treating the two $F=1,2$ sub-ensembles independently. The frequency dependence of the polarizability contributions and expected spontaneous decay coefficients, together with experimental data, are shown in Fig.~\ref{fig:alpha_pol}.

The scalar polarizabilities $(k=0)$ do not affect the polarization of a light beam. The higher order terms are linked to spin-dependent circular $(k=1)$ and linear $(k=2)$ birefringence, named Faraday and Voigt effect, respectively. We assume a quasi one-dimensional scenario with cross section $A$, and describe a coherent laser beam, polarized at $45\degree$, by photon flux $S_y$. Stokes operators $\hat{S}_{x,z}\approx 0$ quantify quantum mechanical uncertainty of the input beam's polarization, see Eq.~\ref{eq:Stokesops} for definitions. For small optical phase shifts $(\ll 1~\mathrm{rad})$, neglecting light retardation and back action onto the traversed atomic ensemble, the polarization rotation and ellipticity of the output beam are measured by the operators
\begin{align}
\hat{S}_x'&=\hat{S}_x-
g_{F}^{(1)} S_y \sum_i\hat{F}_{z,i}\\
\hat{S}_z'&=\hat{S}_z+
g_{F}^{(2)} S_y\sum_i\left(\hat{F}_{x,i}^2-\hat{F}_{y,i}^2\right),
\label{eq:ellipticity}
\end{align}
which sum individual atomic spin operators, with coupling constants  $g^{(k)}_{F}=\alpha^{(k)}_{F}\omega_L/(A\epsilon_0 c\hbar^{k})$, see~\ref{eq:birefringence}.

For known spin states, both signals can in principle be used to measure atom numbers. If all $n_F$ atoms in the $F$-manifold are in the same state, we can express the expectation values by individual atomic operators as
\begin{align}
\Braket{\hat{S}_x'}&=-
g_{F}^{(1)} S_y n_F \Braket{\hat{F}_z},\\
\Braket{\hat{S}_z'}&=
g_{F}^{(2)} S_y n_F \Braket{\hat{F}_x^2-\hat{F}_y^2}.
\end{align}

For standard clock states, which have one zero spin component ($m=0$), the population cannot be detected by measuring Faraday rotation due to lack of any orientation, i.e., $\Braket{\hat{\mathbf{F}}}=0$. But for atoms in an eigenstate of the $\hat{F}_y$-operator, linear birefringence is proportional to $\xi_F(F_y)=\bra{F,F_y}\hat{F}_x^2-\hat{F}_y^2\ket{F,F_y}/\hbar^2$. The moment $\xi_F(m)=(F(F+1)-3m^2)/2$ is extremal for $m=\pm F$ as well as for $m=0$ (bosons) or $m=\pm 1/2$ (fermions). Intermediate Zeeman substates exhibit smaller linear birefringence, which becomes exactly zero only in rare cases including $\ket{0,0},\ket{1/2,\pm1/2}$, $\ket{3,\pm2},\ket{25/2,\pm15/2}, \ket{48,\pm28}, \ket{361/2,\pm209/2}$ etc.

\subsection{\label{sec:citeref} Adiabatic radio-frequency dressing}

In this section, we outline the principle of adiabatic RF dressing and discuss its effect on the measurements of atomic observables.

The magnetic fields that we use in our experiments are generally weak enough to neglect second order Zeeman splitting within each hyperfine manifold. In this case, RF dressing can simply be described as a rotation of an effective magnetic field  $\mathbf{B}_{\mathrm{eff}}$ that combines the effects of real fields and fictitious forces in a rotating frame. For slow enough rotation of this effective field with respect to the rotating frame, the atomic spin will adiabatically follow and precess about the direction of the effective field with constant spin projection along that direction. 

To first order, the time-dependent interaction Hamiltonian of an atom with spin $\mathbf{F}$ of constant magnitude in a magnetic field with static and oscillatory components is given by
\begin{equation}
	\hat{H}=\frac{\mu_Bg_F}{\hbar} \hat{\mathbf{F}}\cdot\left(\mathbf{B}_{\mathrm{RF}}(\omega t)+\mathbf{B}_{\mathrm{DC}}\right),
\end{equation}
where $\mu_B$ is the Bohr magneton and $g_F$ is the Land\'e factor. The oscillating part can best be expressed in terms of spherical polarization components. Choosing $\mathbf{B}_{\mathrm{DC}}=B_{\mathrm{DC}} \mathbf{e}_z$ and using the spherical basis $\mathbf{e}_{\pm}=(\mathbf{e}_x\pm i\mathbf{e}_y)/\sqrt{2}$ and $\mathbf{e}_{\pi}=\mathbf{e}_z$, we can write
\begin{equation}
	\mathbf{B}_{\mathrm{RF}}(\omega t)=\mathrm{Re}\left[(B_+\mathbf{e}_{+}+B_-\mathbf{e}_{-}+B_{\pi}\mathbf{e}_{\pi} )e^{-i\omega t}\right].
\end{equation}
Using corresponding spin components with the conventional normalization of raising and lowering operators  $\hat{F}_{\pm}=\hat{F}_x \pm\hat{F}_y$, the Hamiltonian is expressed as
\begin{align}
 	\nonumber\hat{H}=\frac{\mu_B g_F}{2\hbar}&\left[\left(\frac{B_+}{\sqrt{2}}\hat{F}_+ +\frac{B_-}{\sqrt{2}}\hat{F}_- + B_{\pi}\hat{F}_z \right)e^{-i\omega t}\right.\\
	&\left.+B_{\mathrm{DC}}\hat{F}_z\vphantom{[\left(\frac{B}{\sqrt{2}}\right)}\right]+h.c.
\end{align}
We transform to a frame rotating about the $z$-axis at frequency $\omega$ with a given phase $\varphi$, such that $\hat{H}_{\mathrm{rot}}=\hat{U}\hat{H}\hat{U}^{-1}+i\hbar\frac{\partial}{\partial t}\hat{U}\hat{U}^{-1}$, using the unitary transformation
\begin{equation}
    \hat{U}_{\pm}(t)=e^{i(\pm\omega t+\varphi)\hat{F}_z/\hbar},    
\end{equation}
where the sign of frequency is chosen equal to the sign of the Land\'e factor $g_F$, which determines the sense of rotation that is required to dress atoms resonantly.
Using the identity $e^{\alpha\hat{F}_z}\hat{F}_{\pm}e^{-\alpha\hat{F}_z}=e^{\pm\alpha} \hat{F}_{\pm}$, the rotating frame Hamiltonian becomes
\begin{align}
 	\hat{H}_{\mathrm{rot}}^{\pm}=\frac{\mu_B g_F}{2\hbar}&\left[\frac{B_{\mp}e^{\mp i\varphi}}{\sqrt{2}}\hat{F}_{\mp} e^{-2i\omega t} + B_{\pi}\hat{F}_z e^{-i\omega t}\right.\\
	\nonumber&\left.\hphantom{\left[\right.}+\frac{B_{\pm}e^{\pm i\varphi}}{\sqrt{2}}\hat{F}_{\pm}+\left(B_{\mathrm{DC}}-B_{\mathrm{res}}\right)\hat{F}_z\vphantom{\frac{B_+e^{i\varphi}}{\sqrt{2}}}\right]+h.c.,
\end{align}
where we introduced the positive, resonant field $B_{\mathrm{res}}=\pm\hbar\omega/\mu_B g_F$. 

If the RF field is polarized purely in the $\mathbf{e}_{\pm}$ direction that corresponds to the Larmor precession, i.e., $B_{\mp}=B_{\pi}=0$, atoms will exhibit the same behavior as in an apparently static, effective field
\begin{equation}
    \mathbf{B}_{\mathrm{eff}}^{\pm}=\frac{1}{2} (B_{\pm}e^{\pm i\varphi} \mathbf{e}_{\pm} + c.c.) + (B_{\mathrm{DC}}-B_{\mathrm{res}}) \mathbf{e}_z,
\end{equation}
described by the corresponding effective, rotating frame Hamiltonian
\begin{equation}
\hat{H}_{\mathrm{eff}}^{\pm}=\frac{\mu_Bg_F}{\hbar}\hat{\mathbf{F}}\cdot\mathbf{B}_{\mathrm{eff}}^{\pm}.
\end{equation}
In particular, an atomic spin will adiabatically follow the effective field's orientation provided that any reorientation with $\dot{\mathbf{B}}_{\mathrm{eff,\perp}}=\mathbf{\Omega}\times\mathbf{B}_{\mathrm{eff}}$ occurs at a rate that is much slower than the effective Larmor frequency, i.e., for $|\mathbf{\Omega}|\ll\mu_Bg_F|\mathbf{B}_{\mathrm{eff}}|/\hbar$. When other RF-polarization components are present, the rotating wave approximation can be used, thus neglecting the fast oscillating terms of $\hat{H}_{\mathrm{rot}}$ as long as $\omega\gg\mu_Bg_F|\mathbf{B}_{\mathrm{eff}}|/\hbar$~\cite{Cohen-Tannoudji1998}. 
The resulting behaviour remains the same apart from second order energy shifts~\cite{Bloch1940}. 

We now consider the specific transformation of atomic spin operators when an eigenstate of $\hat{F}_z$, initially prepared in a static field along the $z$-direction, is dressed by adiabatically changing the components of the effective field. For the purpose of this paper, we assume an RF field that is linearly polarized in the $x,y$-plane, described by
\begin{equation}
	\mathbf{B}_{\mathrm{RF}}(\omega t)=B_{\mathrm{RF}}\cos{\omega t}\cdot\left(\mathbf{e}_{x}\cos{\varphi}+\mathbf{e}_{y}\sin{\\\varphi}\right),
\end{equation}
for which case $B_{\pm}=B_\mathrm{RF}e^{\mp i\varphi}/\sqrt{2}$ and $B_{\pi}=0$. The phase $\varphi$ describes the direction of field oscillation and determined our above choice of phase for the rotating frame. Consequently, we find for either effective field
\begin{equation}
    \mathbf{B}_{\mathrm{eff}}^{\pm}=\frac{B_{\mathrm{RF}}}{2}\mathbf{e}_x + (B_{\mathrm{DC}}-B_{\mathrm{res}}) \mathbf{e}_z.
\end{equation}
The initial state $\left|F, F_z\right\rangle$ is an eigenstate of $\hat{U}_{\pm}(t)$ and appears identical in both laboratory and rotating frame. For $B_{\mathrm{RF}}=0$ it is an eigenstate of either frame's Hamiltonian and differs only in its evolution of dynamical phase or its quasi-energy, which we can ignore for our purposes. Upon changing the effective field, we obtain the adiabatic state by applying the corresponding rotation about the (rotating) $y$-axis according to
\begin{equation}
    \left|\Psi_{\mathrm{rot}}\right\rangle=e^{i\theta\hat{F}_y}\left|F,F_z\right\rangle,
\end{equation}
by an angle
\begin{equation}
\theta=\frac{\pi}{2}-\mathrm{tan}^{-1}{\frac{B_\mathrm{DC}-B_{\mathrm{res}}}{B_{\mathrm{RF}}/2}}.
\end{equation}
The same state in the laboratory frame is then given by
\begin{equation}
\left|\Psi(t)\right\rangle=\hat{U}_{\pm}^{-1}(t)\left|\Psi_{\mathrm{rot}}\right\rangle.
\end{equation}
Finally, we can express any laboratory frame atomic observable $\hat{O}$ using
\begin{align}
    \left\langle\Psi(t)\right|\hat{O}\left|\Psi(t)\right\rangle&=\left\langle F,F_z\right|\hat{R}_{\pm}\hat{O}\hat{R}_{\pm}^{-1}\left|F,F_z\right\rangle,\\
    \hat{R}_{\pm}(t)&=e^{-i\theta\hat{F}_y}\hat{U}_{\pm}(t).
\end{align}

The result is a time-dependent geometrical rotation of coordinates given by the explicit transformation 
\begin{align}\label{eq:rot}
    &\hat{\mathbf{F}'}(t)=\hat{R}_{\pm}(t)\hat{\mathbf{F}}\hat{R}_{\pm}^{-1}(t)=\mathbf{R}_{\pm}(t)\hat{\mathbf{F}}\\
    \nonumber&=\left(
    \begin{array}{ccc}
        \cos{\theta}\cos{\phi_{\pm}(t)} &-\sin{\phi_{\pm}(t)} &-\sin{\theta}\cos{\phi_{\pm}(t)}\\
        \cos{\theta}\sin{\phi_{\pm}(t)} &\cos{\phi_{\pm}(t)} &-\sin{\theta}\sin{\phi_{\pm}(t)}\\
        \sin{\theta} &0 &\cos{\theta} 
    \end{array}
    \right)
    \hat{\mathbf{F}},
\end{align}
where we defined $\mathbf{R}_{\pm}(t)=\mathbf{R}_z(\phi_{\pm}(t))\mathbf{R}_y(-\theta)$ as a combination of rotations about coordinate axes according to $\mathbf{R}_k\mathbf{v}=\mathbf{e}_k(\mathbf{e}_k\cdot\mathbf{v})(1-\cos{\alpha})+\mathbf{v}\cos{\alpha}+\mathbf{e}_k\times\mathbf{v}\sin{\alpha}$ (Rodrigues' rotation formula) and the time dependent angle $\phi_{\pm}(t)=\pm\omega t+\varphi$.

\subsection{Linear birefringence of dressed eigenstates\label{subsec:dressed_eigenstates}}

To consider different experimental geometries, in particular for light propagation parallel or orthogonal to the static field, we use rotated light coordinates, expressed by a general rotation matrix $\mathbf{M}$, such that $(x',y',z')^T=\mathbf{M}(x,y,z)^T$. 
Using Eq.~\ref{eq:ellipticity}, linear birefringence of eigenstates of the dressed Hamiltonian is then measured by
\begin{align}
    \nonumber\hat{S}_{z'}(t)&=\hat{S}_{z'}(t)+g_F^{(2)}S_y\sum_{i=1}^{n_F} \left[\hat{F'}_{y',i}^2(t)-\hat{F'}_{x',i}^2(t)\right]\\
    &=\hat{S}_{z'}(t)+g_F^{(2)}S_y\sum_{i=1}^{n_F}\hat{\mathbf{F}}_i^T\mathbf{Q}_{\pm}\hat{\mathbf{F}}_i,
\end{align}
introducing the quadratic form
\begin{equation}
    \mathbf{Q}_{\pm}=\mathbf{R}_{\pm}^T(t)\mathbf{M}^T
    \left(
        \begin{array}{ccc}
             -1& 0&0\\
             0&1&0\\
             0& 0&0 
        \end{array}
    \right)
    \mathbf{M}\mathbf{R}_{\pm}(t).
\end{equation}

Since the matrix $\mathbf{Q}_{\pm}$ is symmetric and the expectation values of mixed anti-commutators vanish for the original state $\bra{F,F_z}\{\hat{F}_j,\hat{F}_k\}\ket{F,F_z}_{j\neq k}=0$, the expression for the expected signal from an ensemble of identically prepared atoms reduces to the trace
\begin{equation}
    \Braket{\hat{S}_{z'}(t)}=g_F^{(2)}S_y n_F\sum_{j=1}^3\Bra{F,F_z} Q_{\pm}^{j,j}\hat{F}_j^2\Ket{F,F_z},
\end{equation}
which can be expressed in terms of spectral RF components as
\begin{equation}
\Braket{\hat{S'}_z(t)}=g_F^{(2)}S_yn_F\frac{\xi_F(F_z)\hbar^2}{2}\sum_{n=0}^2 h_n(\theta)e^{in\omega t}+c.c.
\label{eqn:spectraloutput}
\end{equation}

We can restrict the description of light geometry to two degrees of freedom, because rotations about the laboratory fixed $z$-axis are equivalent to a rotated RF field, already described by $\varphi$.
We choose sequential rotations $\mathbf{M}=\mathbf{R}_x(\alpha)\mathbf{R}_{y}(\beta)=\mathbf{R}_{y'}(\beta)\mathbf{R}_{x}(\alpha)$ leading to the result
\begin{align}
    \label{eqn:full}
    &\left(
    h_0,h_1,h_2\right)^T(\theta)=\\
    \nonumber&\left(\begin{array}{l}
        \frac{1+3\cos{2\theta}}{4}
        \left(\frac{\cos^2\!{\beta}}{2}{\scriptstyle -}\frac{(3-\cos{2\beta})\cos{2\alpha}}{4}
        \right)\\
        {\scriptstyle \sin{2\theta}}
        \left(\frac{\cos{\alpha}\sin{2\beta}}{2}{\scriptstyle \mp i} \frac{(3-\cos{2\beta})\sin{2\alpha}}{4}
        \right)e^{\pm i\varphi}\\
        {\scriptstyle -\sin^2{\theta}}
        \left(\frac{(3-\cos{2\beta})\cos^2\!{\alpha}+2\cos{2\beta}}{4}{\scriptstyle \mp i} \frac{\sin{\alpha}\sin{2\beta}}{2}
        \right)e^{\pm 2i\varphi}\\
    \end{array}\right).
\end{align}

For the parallel setting $\alpha=\beta=0$, the chosen coordinate systems for atomic and light variables coincide. In this case, the spectral components reduce to
\begin{align}
    \left(\begin{array}{l}
        h_0\\h_1\\h_2\\
    \end{array}\right)_{\parallel}\!(\theta)
    =
    \left(\begin{array}{l}
        0\\
        0\\
        -e^{\pm 2i\varphi}\sin^2\!{\theta}\\
    \end{array}\right).
    \label{eqn:orthogonal}
\end{align}

\begin{figure}[t]
\includegraphics[width=0.45\textwidth]{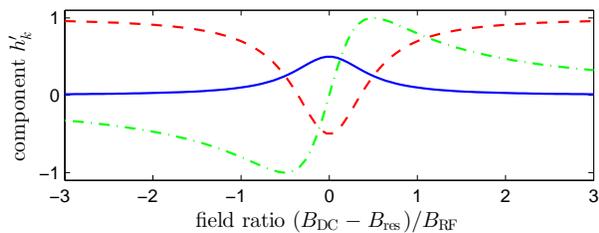}
\caption{Principal behaviour of zeroth (dashed red, $n=0$), first (dash-dotted green, $n=1$) and second (solid blue, $n=2$) harmonic signal components across RF resonance for the orthogonal case ($\beta=\pi/2$) plotted as $h'_n=(\pm i)^n\sqrt{2}h_n$ with $\varphi=0$, $\alpha=\pi/4$.}
\label{fig:harmonics}
\end{figure}

A setting with light propagation orthogonal to the static field is described by
$\beta=\pi/2$. In this case, $\alpha$ describes a rotation of beam polarization, with $\alpha=0$ for unchanged polarization, i.e., at $45^{\circ}$ with respect to the static field. 
The amplitudes of the spectral components are then given by
\begin{align}
    \left(\begin{array}{l}
        h_0\\h_1\\h_2\\
    \end{array}\right)_{\perp}\!(\theta)
    =
    \left(\begin{array}{l}
        -\frac{1}{4}\left(1+3\cos{2\theta}\right)\cos{2\alpha}\\
        \mp i e^{\pm i\varphi}\sin 2\theta \sin{2\alpha}\\
        -\frac{1}{2}e^{\pm 2i\varphi}\sin^2\!{\theta}\cos{2\alpha}\\
    \end{array}\right).
    \label{eqn:parallel}
\end{align}
The principal behaviour of these functions across RF resonance is shown in Fig.~\ref{fig:harmonics}.
\noindent

The results show that due to the axial symmetries of both the setup and the initial state, a parallel measurement only produces signals at the second harmonic, i.e., at frequency $2\omega$. This setting also leads to the maximum possible signal oscillation with full swing between $\pm S_{\mathrm{max}}=\pm g_F^{(2)}S_yn_F\xi_F(m)\hbar^2$ when the RF resonance condition $\theta=0$ is met. The orthogonal setting with $\alpha=0$ contains a DC part that is reminiscent of undressed detection with off-resonant amplitude $S_{\mathrm{max}}$ and leads to a weaker signal at $2\omega$ on resonance with an amplitude swing between $0$ and $-S_{\mathrm{max}}$. In both cases, a signal at frequency $\omega$ arises only due to misalignment or rotated light polarization, with a zero crossing at resonance.

For detection of atomic population, the variations in signal strength will become important. Both, changes in magnitude of the static field $B_{\mathrm{DC}}$, which shifts the resonance condition, as well as field rotations or equivalent beam misalignment affect the resonant $2\omega$-signal only to second order. Since the RF amplitude has no effect, it is advantageous to use higher RF amplitudes to broaden the resonance. The signal becomes less sensitive to fluctuations of external magnetic fields reducing the requirements for magnetic field shielding. A limit to this strategy will be imposed by effects from second order Zeeman splitting, which we do not analyze here.   

\section{Experimental realisation\label{sec:Experiment}}

\subsection{State preparation\label{subsec:state_preparation}}

\begin{figure}[b]
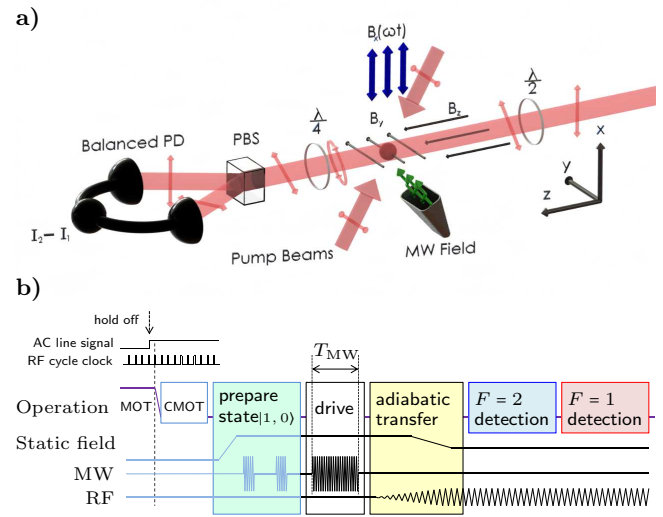

\centering
\begin{lpic}[]{Fig4a_setup(8.5cm)}
    \lbl[bl]{0,45,0;\bf{a)}}
\end{lpic}
\begin{lpic}[]{Fig4b_sequence(8.5cm)}
    \lbl[bl]{0,80,0;\bf{b)}}
    \lbl[bl]{27,72,0;\tiny{\textsf{hold off}}}
    \lbl[bl]{6,63,0;\tiny{\textsf{AC line signal}}}
    \lbl[bl]{4,57,0;\tiny{\textsf{RF cycle clock}}}
    \lbl[bl]{102.5,60,0;\scriptsize{$T_\mathrm{MW}$}}
    \lbl[bl]{0,40,0;\scriptsize{Operation}}
    \lbl[bl]{36,42,0;\tiny{\textsf{MOT}}}
    \lbl[bl]{51,42,0;\tiny{\textsf{CMOT}}}
    \lbl[bl]{70,44,0;\scriptsize{\textsf{prepare}}}
    \lbl[bl]{70,37,0;\scriptsize{\textsf{state}}\tiny{\textsf{$|1,0\rangle$}}}
    \lbl[bl]{103,41,0;\scriptsize{\textsf{drive}}}
    \lbl[bl]{124,44,0;\scriptsize{\textsf{adiabatic}}}
    \lbl[bl]{124,37,0;\scriptsize{\textsf{transfer}}}
    \lbl[bl]{158,44,0;\scriptsize{$F=2$}}
    \lbl[bl]{158,37,0;\scriptsize{\textsf{detection}}}
    \lbl[bl]{190,44,0;\scriptsize{$F=1$}}
    \lbl[bl]{190,37,0;\scriptsize{\textsf{detection}}}
    \lbl[bl]{0,29,0;\scriptsize{Static field}}
    \lbl[bl]{20,18,0;\scriptsize{MW}}
    \lbl[bl]{24,10,0;\scriptsize{RF}}
\end{lpic}
\caption{Experimental setup. a) A laser cooled rubidium sample is prepared in a superposition of two clock states by $\pi$-polarized optical pumping and microwave driving in a static field along $y$. After adiabatic dressing with a magnetic RF field along $x$ and optional rotation of the static field into the $z$-direction, linear birefringence of the sample is probed polarimetrically by two consecutive laser pulses propagating along $z$. b) Main timing elements of the experimental procedure.}
\label{fig:ExpSetup}
\end{figure}

We apply our detection method to an ensemble of approximately $10^8$ $^{87}$Rb atoms, which we prepare in superpositions of the two clock states $\ket{F=1, m_F=0}$ and $\ket{F=2, m_F=0}$ by driving the clock transition with a resonant microwave pulse of variable duration. 

A sketch of the experimental setup is shown in Fig.~\ref{fig:ExpSetup}~a). In order to start from a pure state, we use an optical pumping and cleaning sequence to initially prepare atoms in $\ket{F=1, m_F=0}$.
After releasing a cloud of atoms from a standard,  transiently compressed magneto-optical trap~\cite{Petrich1994}, we perform optical molasses cooling while we ramp up a weak magnetic field in the $y$-direction to $\approx 0.3~\mathrm{G}$. We then replace the standard $F=1\to F'=2$ repumping beam by a pair of counterpropagating, $\pi$-polarized beams tuned near the $F=1\to F'=1$ transition on the $ \mathrm{D}_1$ line for optical pumping. We use an intensity of $80~\mathrm{\mu W cm^{-1}}$ and a red detuning of $-30~\mathrm{MHz}$ to reduce re-absorption of scattered photons. This method continues to provide cooling and avoids directional forces while atoms accumulate in the now dark $\ket{1, 0}$ state. After a period of $6~\mathrm{ms}$ and sequential switch-off of first pump then cooling beams, we achieve $(70\pm5)\%$ population in $\ket{1,0}$ with a final temperature of $(80\pm10)~\mathrm{\mu K}$ and the remaining atoms populating the $\ket{1, \pm1}$ states. Purification of the state is achieved by coherent transfer of atoms from $\ket{1,0}$ to $\ket{2,0}$ using a resonant microwave $\pi$-pulse emitted from a sawed-off waveguide and a raised magnetic field of $\approx 0.5~\mathrm{G}$, followed by a short pulse from the original repumping beam and a second microwave $\pi$-pulse, converting $\ket{2, 0}$ back to $\ket{1, 0}$. Incoherently transferred atoms then populate only $F=2$ levels. We push these away from the cloud by shining a single resonant beam on the cycling $F=2\to F'=3$ transition on the $\mathrm{D}_2$-line, leaving only the purified $\ket{F=1, m_F=0}$ state.

\subsection{Dressed state detection\label{subsec:Dressed_state_det}}

Figure~\ref{fig:ExpSetup} b) shows the experimental sequence for state preparation, dressing and state detection. While the purified ensemble is in free fall, we apply a resonant microwave pulse for a variable duration $T_{\mathrm{MW}}$ to drive high-contrast Rabi cycles and prepare superpositions of the two clock states. Atoms are then adiabatically dressed with a magnetic RF field in the $x$-direction with frequency $\omega = 2\pi \times 180~\mathrm{kHz}$, generated by an external resonant coil. The RF field amplitude is ramped up to $\approx 15~\mathrm{mG}$ over $4~\mathrm{ms}$ while the static magnetic field is ramped down to a magnitude of $B_{\mathrm{DC}} \approx 260~\mathrm{mG}$, which tunes the atomic Larmor frequency near resonance. For most experiments, the static field is simultaneously rotated from the $y$-direction into the $z$-direction. This procedure maintains the magnitude of the total collective spin as well as its alignment with the effective field such that the initial atomic spin projection $F_y=0$ then rotates within the $x,y$-plane. While the total populations within each $F$-manifold remain unchanged, the atomic state then obeys $F_x\cos(\omega t)\pm F_y \sin(\omega t)=0$, where the sign of rotation depends on the state-dependent Land\'e factor $g_F$. 

\begin{figure}[t]
\centering
\includegraphics[width=0.48\textwidth]{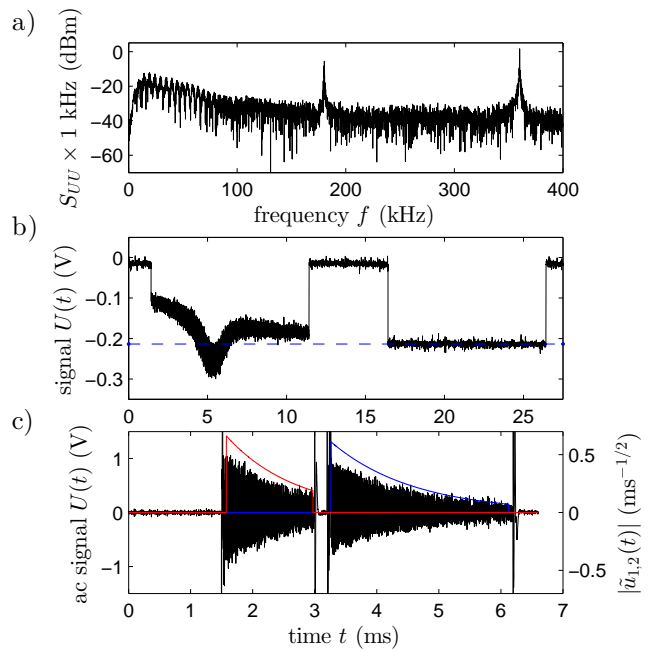}
\caption{Typical experimental signals. a) Single-sided, power spectral density of the amplified, high-pass filtered signal. Atomic signals arise at $\omega$ and $2\omega$. b) Direct signal for $F=2$, recorded during a sweep of static field strength across the RF resonance in the orthogonal setting. Atoms are removed before a second probe pulse is used to determine the signal offset from imperfect detector balance (dashed line). The signal matches the theoretical response but shows probe induced decay. c) Amplified signal for two-color measurement of both state populations in the parallel setting. Envelopes of the used temporal mode functions $\tilde{u}_{1,2}(t)$ are shown in red (first pulse, $F=2$) and blue (second pulse, $F=1$).}
\label{fig:SampleTraces}
\end{figure}

We use two-color detection to distinguish populations in the two hyperfine manifolds. The two $D_1$-line optical frequencies are detuned by $-400~\mathrm{MHz}$ from the $F=2\to F'=2$ transition and by $+240~\mathrm{MHz}$ from the $F=1\to F'=1$ transition, respectively, avoiding two-photon resonance. Due to their separation of $\approx 6.7~\mathrm{GHz}$, the interaction of each field with the atomic cloud is dominated by population in one of the two hyperfine states. The beams have perpendicular, linear polarizations and are combined with a Wollaston prism to co-propagate through the atomic ensemble. A half-wave plate allows us to co-rotate the planes of polarization with respect to the coordinate axes, typically adjusted to $\pm45\degree$ orientation. To observe the adiabatic dressing process, the detection beams can be active during the magnetic field ramps. For state detection, we let the magnetic field amplitudes reach constant values before the two lasers are pulsed either consecutively or, in some circumstances, simultaneously as detailed below.
Upon interaction with the atoms, the beams become elliptically polarized, where the ellipticity or phase shift between $\pi$- and $\sigma$-polarized components is proportional to the atomic density in the respective states. The phase shift is measured polarimetrically with a circular analyzer comprising a quarter-wave plate, a Wollaston prism, a balanced photodetector pair (Thorlabs PDB210A) and an optional high-pass filtering RF amplifier (Minicircuits Model ZFL-1000+). The quarter-wave plate is aligned such that the differential photo current measures the difference between right- and left-hand circularly polarized components. The output voltage $U$ is proportional to the observed ellipticity, i.e., $U(t)=g_{\mathrm{el}}S_z(t)$ with electronic gain $g_{\mathrm{el}}$. Figure~\ref{fig:SampleTraces} shows examples of typical raw detector signals together with a signal spectrum, which shows that signals arise at $180~\mathrm{kHz}$ and $360~\mathrm{kHz}$ above a noise floor that is limited by photon shot noise at frequencies above $\approx 150~\mathrm{kHz}$. At lower frequencies, the spectrum is dominated by (ac-filtered) square-pulse transients from imperfectly balanced detector signals. As expected, the main contribution to the RF signal is found at frequency $2\omega$. We also detect signals at frequency $\omega$ in case of geometric misalignment.

The raw signals are processed via digital lock-in detection. As can be seen in Fig.~\ref{fig:SampleTraces}~c), the atomic signals decay due to spontaneous emission induced by the probe beams. We obtain signal values proportional to state populations by extracting spectral mode amplitudes $m_{1,2}=\int\!u_{1,2}^*(t)U(t)dt$ from the RF signal $U(t)$ with $L^2$-normalized temporal mode functions $u_{1,2}(t)=\tilde{u}(t)e^{2i(\omega t\pm\varphi)}$. In the case of square laser pulses of duration $T$, their envelopes take the form
\begin{equation}
\label{eq:modefunction}
\tilde{u}(t)=
    \begin{cases}
    \sqrt{\frac{2\gamma}{1-e^{-2\gamma T}}}e^{-\gamma t}, & \mathrm{if}\ 0\leq t\leq T \\
      0, & \mathrm{otherwise},
    \end{cases}
\end{equation}
with experimentally determined, probe power dependent decay rates $\gamma$. For higher probe powers, we use shaped pulses to avoid light-shift induced excitation of Larmor precession about the effective field, which may occur at Rabi frequency $\Omega_{\mathrm{RF}}$ corresponding to the RF field amplitude. For shaped pulses, we use heuristically adapted mode-functions (see shaped pulses in Fig.~\ref{fig:singlepulse}). Our frequencies and mode functions allow for slowly varying envelope approximations with negligible spectral overlap of signals from different harmonics.

The mode amplitudes are referenced to the input light according to $m'_{1,2}=m_{1,2}/P_{1,2}$, where $P_{1,2}$ are the simultaneously measured, pulse-averaged probe powers, proportional to $S_y$. This is used to correct for small light power fluctuations, neglecting power-dependent changes in the decay rates. We extract real signals by correcting each mode amplitude for an experimentally determined constant phase $\varphi_{1,2}$, which includes effects from the geometry and choice of polarization, see Eq.~\ref{eqn:orthogonal} and \ref{eqn:parallel}, as well as phases introduced by the detection electronics. State populations $n_{1,2}$ are estimated from the mode amplitudes, assuming $m'_{1,2}=g^{(\mathrm{exp})}_{1,2}n_{1,2}$, where the experimentally determined signal gains $g^{(\mathrm{exp})}_{1,2}$ are calibrated against atom number estimates from absorption imaging data and account for the combined factors of interaction coefficients,
detuning from resonance, photon energy, and electronic gain. Finally, values for individual measurements of the normalized Bloch-vector component are obtained as $\sigma_z=(n_2-n_1)/(n_2+n_1)$.

\begin{figure}[t]
\centering
\includegraphics[width=0.48\textwidth]{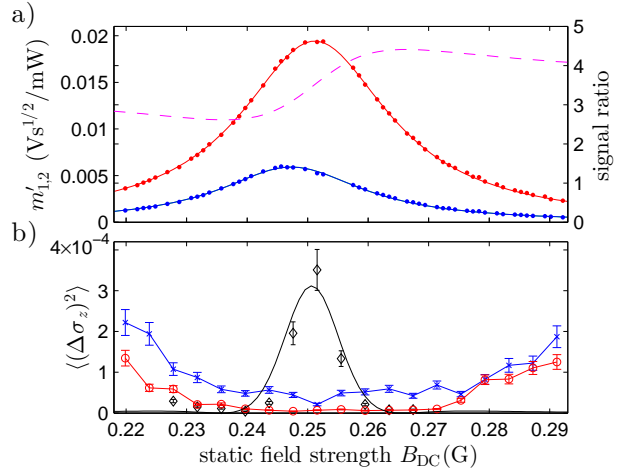}
\caption{Scan across RF resonance. 
a) Experimental mode amplitudes (small circles) at $2\omega$ arising from an equal superposition of the two clock states together with fitted $h_2$ functions (solid lines, left axis) and their ratio (dashed line, right axis) are shown as a function of static field strength in the parallel setting. The stronger signal (upper red curve) arises from atoms in $F=2$ due to larger $\xi_F(0)$. b) Measured variance $\var{\sigma_z}$ as a function of static field amplitude. The noise is reduced (red cicrcles) by synchronizing with an AC mains signal compared to asynchronous measurements (blue crosses). Deliberately introduced static field noise (black diamonds) leads to peaked behaviour following the derivative of signal ratio.}
\label{fig:Scan}
\end{figure}

Figure~\ref{fig:Scan}~a) shows measured mode amplitudes for a scan across RF resonance in the parallel setting and approximately equal populations in the two clock states. The fit of model functions according to the $h_2$-component of Eq.~\ref{eqn:parallel} allows us to extract position and width of the resonances, which we use to calibrate the strength of the applied static field as well as the RF field amplitudes. In this case, the amplitudes were measured to be $|B_+|=(15\pm1~\mathrm{mG})$ and $|B_-|=(14\pm1~\mathrm{mG})$. We attribute the small amplitude difference to stray RF fields from induced eddy currents that make the field polarization slightly elliptic at the location of the atomic ensemble.

\begin{figure}[tb]
\centering
\includegraphics[width=0.45\textwidth]{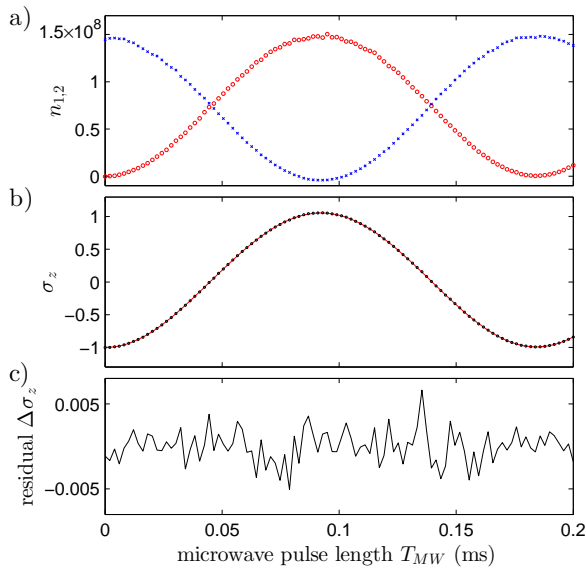}
\caption{Experimental detection of Rabi cycles. a) The populations of both clock states $|1,0\rangle$ (blue crosses) and $|2,0\rangle$ (red circles) are measured as a function of microwave pulse duration. b) Normalized population difference (circles) together with model function (solid line). c) The residuals indicate an oscillating noise amplitude. Each data point corresponds to a single experimental cycle.}
\label{fig:Rabi}
\end{figure}

In order to measure atomic populations with high signal-to-noise ratio it is desirable to measure the corresponding mode amplitudes at $2\omega$ in the parallel setting exactly on RF resonance. Here, the signals are maximal and exhibit only a second order dependence on the static magnetic field, which should reduce noise from external field fluctuations. However, it is important to note that even in the regime of vanishing second order Zeeman splitting at low magnetic fields, the hyperfine structure leads to a first order difference of Larmor frequency between the two hyperfine manifolds. The two corresponding $g$-factors differ in magnitude by $\Delta| g_F|=-2g_I$, where $g_I$ is the (negative) nuclear $g$-factor. For $^{87}$Rb this corresponds to a field dependent frequency difference of $-2g_I\mu_B/h = 2.78565~\mathrm{kHz/G}$~\cite{Steck2015}.
This affects the suppression of common mode noise when $\braket{\hat{\sigma}_z}$ is estimated from the two signals measured at the same static field. Common mode noise includes total atom number fluctuations as well as signal strength variations. While the suppression of noise due to external field fluctuations improves with resonance width, it is highest when the ratio of the two signal strengths as a function of external field is extremal or, equivalently, when the ratio of signal strength to signal slope is identical for the two states. The static field that meets this condition depends on frequency shift and widths of both resonances. For equal resonance width, i.e., $|B_+|=|B_-|=|B_{\mathrm{RF}}|/\sqrt{2}$ and resonant field difference $\Delta B=-\hbar\omega/2g_I\mu_B$, the optimal static field is found shifted from the resonance mean by $\pm\frac{1}{2}\sqrt{|B_{\mathrm{RF}}|^2+(\Delta B)^2}$. As a consequence, signal strength must be traded for maximal common mode noise suppression. This can be improved by deliberately increasing the imbalance between the two $B_{\pm}$-components, which shifts the optimal point closer to the resonance peaks. 

Figure~\ref{fig:Scan}~b) shows experimental variances $\var{\sigma_z}$ for an equal superposition of clock states across the resonance for different experimental conditions. Away from resonance, noise increases due to diminishing signal strength at constant detection noise (electronic and photon shot noise). For a small amount of deliberately introduced noise in the static field amplitude, the resulting variance peaks near maximum signal strength and follows the expected behaviour. Generally, we achieve best performance near the point of stationary signal ratio closest to signal maximum. In our unshielded experiment, disabling synchronization with a 50~Hz line signal increases noise, which we attribute mainly to state preparation noise in the fluctuating environment as this noise contribution remains fairly constant across the scan.

\begin{figure}[tb]
\centering
\includegraphics[width=0.45\textwidth]{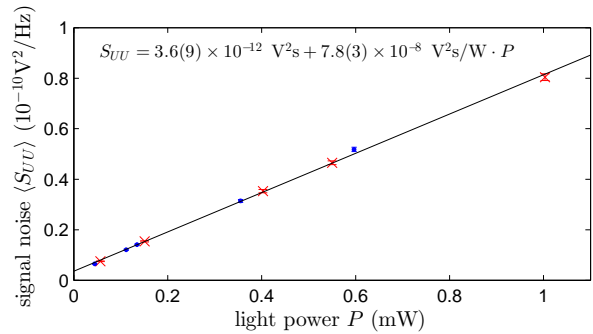}
\caption{Scaling of detection noise with probe power. The linear dependence confirms shot noise limited performance for both probe lasers. Electronic noise is negligible in the typical operating range of a few hundred microwatts probe power. The shot noise scaling can be used to calibrate the electronic gain $g_\mathrm{el}=U/S_z=2\hbar\omega_L U/\Delta P$, relating output voltage $U$ to photon flux or light power difference $\Delta P$ incident on the two detectors. 
For pure shot noise from the input light field of power $P$ and known quantum efficiency of the detector $\eta<1$ (electrons per photon), the electronic gain can be measured as 
$g_\mathrm{el}=2\sqrt{\eta\hbar\omega_L \zeta \braket{S_{UU}}/P}\approx1.3\times10^{-13}\mathrm{V/Hz}$, assuming quantum efficieny $\eta= 0.86$ and an estimated noise power correction factor $\zeta\approx0.5$ due to aliasing.}
\label{fig:lightnoise}
\end{figure}

Measurement results of relative population difference $\sigma_z$ for driven Rabi cycles are shown in Fig.~\ref{fig:Rabi}. We observe high contrast fringes of Rabi frequency $\Omega_{\mathrm{RF}}=5.5~\mathrm{kHz}$, which we model including a small exponential decay accounting for in-homogeneous microwave coupling across the atomic cloud. The residuals typically show noise variances on the order of $10^{-6}$ to $10^{-5}$ varying across individual Rabi cycles. The noise is usually somewhat larger in the vicinity of zero crossings  and increases with the number of cycles, indicating a contribution of state-preparation noise that scales with duration and power of the microwave driving. 

\subsection{Noise analysis \label{subsec:noise_scaling}}

\begin{figure}[tb]
\centering
\includegraphics[width=0.45\textwidth]{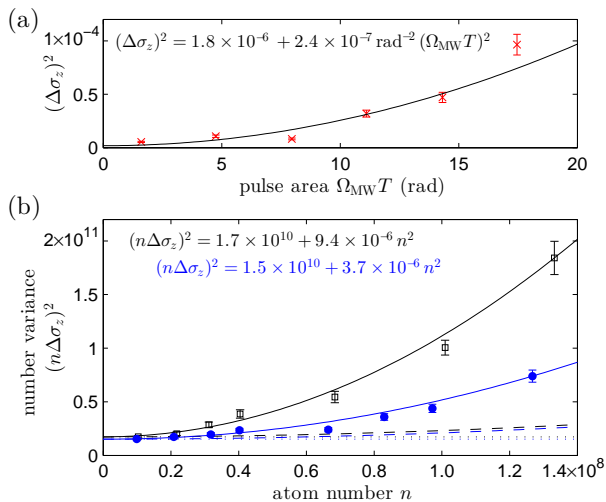}
\caption{Analysis of technical noise for (anti-)symmetric superpositions of the two clock states. a) State preparation noise is identified by driving the clock transition with odd multiples of $\pi/2$ pulse areas and quantifying the quadratic scaling of variance for 100 measurements. For a $\pi/2$-pulse, we estimate an uncertainty contribution of $\Delta\sigma_z=\sqrt{2.4\times10^{-7}}\cdot\pi/2\approx0.08\%$. b) Experimental data for variance of atom number difference show quadratic scaling with total atom number $n$ for two experimental conditions. Signals were measured at constant magnetic field, fulfilling the RF resonance condition only for atoms with $F=2$ (black squares). A small magnetic field shift introduced between the two probe pulses allows for resonant measurements on both states and reduces technical noise introduced by field fluctuations (blue circles). The model fits (solid lines) separate photon shot noise equivalent (dotted) and technical noise. Dashed lines indicate the estimated level of state preparation noise above photon shot noise.}
\label{fig:noiseplot}
\end{figure}

Analysis of different noise contributions to our measurements and distinction between technical and quantum noise can be based on parameter scaling.

Our balanced detector pair (Thorlabs PDB210A) is photon shot noise limited, confirmed by the linear dependence of noise power spectral density $S_{UU}$ of the RF signal on light power in absence of atoms, see Fig.~\ref{fig:lightnoise}.
The detection electronics, including amplification and analog-to-digital conversion, introduce a small amount of electronic noise that is negligible for the used light powers of typically a few hundred microwatts. In principle, the shot noise scaling allows for the determination of electronic gain $g_\mathrm{el}$ and thus linking the output amplitude to observed atom number $n_F$, according to Eq.~\ref{eqn:spectraloutput}. In practice, separate calibration is required due to inhomogeneous atomic densities and Gaussian beam profiles. The measurements presented here also suffer from the lack of a dedicated anti-aliasing filter, which leads to increased noise in the observed RF frequency band due to aliasing of photon shot noise. 

Further noise stems from fluctuations in signal strength, caused by magnetic field fluctuations, as well as varying laser detunings, beam steering and imperfect correction of light power fluctuations. Additional technical noise stems from the microwave driving and thus preparation of the atomic state, and ultimately atomic shot noise. For further analysis, we generated (anti\-/)symmetric superpositions of the two clock states, i.e., $\braket{\hat{\sigma}_z}=0$ and measured the variance of relative population difference for different atom numbers and microwave durations, see Fig.~\ref{fig:noiseplot}. The scaling with atom number shows that measurements at low atom number are limited by photon shot noise while measurements at high atom number ($n\approx10^8$) are dominated by technical noise contributions on the order of $\langle(\Delta\sigma_z)^2\rangle=10^{-6}-10^{-5}$, depending on the precise setting of static magnetic field strength. 
Our photon shot noise equivalent atom number resolution is $\Delta n\approx\sqrt{1.5\times10^{10}}\approx1.2\times10^5$, i.e., $\approx22\mathrm{dB}$ above atomic shot noise for $n=10^8$ atoms. The scaling with microwave duration shows a small contribution of state preparation noise.

\subsection{State-to-quadrature mapping\label{subsec:state_quadratures}}

\begin{figure}[tb]
\centering
\includegraphics[width=0.45\textwidth]{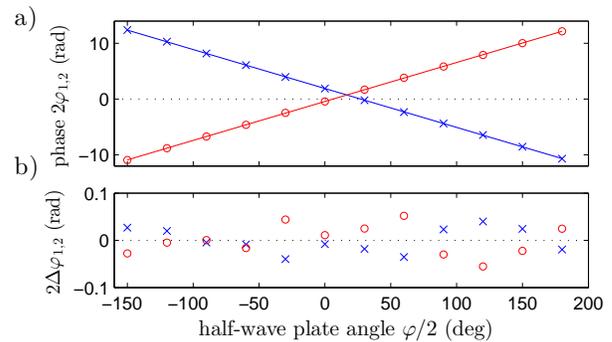}
\caption{Dependence of $2\omega$-signal phases $2\varphi_{1,2}$ on half-wave plate angle in the parallel setting (corresponding to $\varphi/2$). a) Experimental data are shown together with linear fits with slopes $\pm(3.99\pm0.01)~\mathrm{deg/deg}$, matching the expected value of $\pm4$. b) The strong anti-correlation of residuals shows that the dominant uncertainty stems from the wave-plate setting. }
\label{fig:f1f2phases}
\end{figure}

We attribute a significant amount of technical noise in our measurements to the use of two independent probe beams, which do not probe the exact same volume. As a consequence, fluctuations in the position and shape of the atomic ensemble will translate into independent signal fluctuations. In addition, the lasers exhibit independent frequency and power fluctuations. While power fluctuations are co-measured and compensated for, small imperfections like non-linearity and electronic noise in the detection system will degrade the performance. It is in principle possible to use a two-colour beam from a single laser, or phase and amplitude locked beams. Simultaneous detection of both states will then achieve suppression of common mode noise. To distinguish the two signals, use can be made of the fact that the RF phase of the detected signal is adjustable and state-dependent, as represented by the sign of the phase $\varphi$ in Eq.~\ref{eqn:full}-\ref{eqn:parallel}.

\begin{figure}[!t]
\centering
\includegraphics[width=0.48\textwidth]{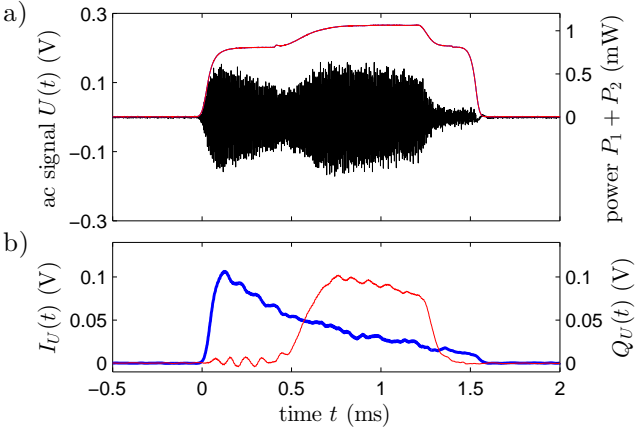}
\caption{Demonstration of state-to-quadrature mapping. Two probe pulses of different duration are sent through the atomic cloud with temporal overlap. a) The IQ-modulated RF signal from the atomic response (oscillating black curve) is shown together with total the light power $P_1+P_2$ (upper red curve). The light pulse edges have been shaped to suppress excitation of Larmor precession in the effective field, which can occur at higher laser powers. b) The RF response is demodulated with $10~\mathrm{kHz}$ bandwidth centred at $2\omega=360~\mathrm{kHz}$ and separated into two orthogonal quadratures $I_U(t)$ and $Q_U(t)$. The light polarization was adjusted to obtain out-of-phase responses from atoms in $F=1$ (thick blue line) and $F=2$ (thin red line), i.e., by choosing $\varphi_2-\varphi_1\approx\pi/2$.}
\label{fig:singlepulse}
\end{figure}

The phase $\varphi$ describes the orientation of the RF field with respect to the coordinate axes. In the parallel setting, a rotation of the RF field is fully equivalent to a rotation of light polarization. We confirmed this by populating only one hyperfine manifold at a time and measuring the RF phase of the $2\omega$-signal for various angles of a half-wave plate that we use to co-rotate the linear input polarization of our light fields. The results presented in Figure~\ref{fig:f1f2phases} exhibit the expected behaviour. The choice of angle therefore allows for direct subtraction of signals at the photodetector for $\Delta(2\varphi)=\pi$ as well as mapping of the two signals onto orthogonal rf quadratures, called IQ-modulation, for $\Delta(2\varphi)=\pm\pi/2$. The latter situation is demonstrated in Fig.~\ref{fig:singlepulse}, which shows experimental data of the two signals for a superposition of clock states.

\section{Conclusions\label{sec:conclusions}}
We analysed and demonstrated dispersive detection of alkali atoms in radio-frequency dressed states. An experimentally simple polarimetric setup allows for low-noise measurements of atom numbers due to modulation of the atomic response at radio-frequencies. 
Linear birefringence measurements of driven Rabi-cycles between atomic clock states show technical noise on the cycle-phase on the order of 2 mrad. Future improvements may include the use of simultaneous probing of both states with two frequencies generated by a single, modulated laser. The ability to perform state to RF-quadrature mapping makes it possible to measure differential state population directly and potentially generate spin squeezing in the regime of strong light-matter coupling at sufficient optical density. The method can be used in various internal state atom interferometry experiments and may be extended to other dressed state schemes.

The datasets generated for this paper are accessible at \cite{openaccessdata}
(Nottingham Research Data Management Repository).

\section{Acknowledgements}
This work was funded by the EU (FP7-ICT-601180) and EPSRC (EP/M013294/1, EP/J015857/1).
SJ was supported by the EU (FP7-PEOPLE-2012-ITN-317485). We gratefully acknowledge useful discussions with I. Lesanovsky and thank K. Poulios for help with final calibrations.

\appendix

\section{Dispersive measurement operators}
\label{sec:dispersiveinteraction}
We consider a quasi one-dimensional situation with cross section $A$, light propagating along $\mathbf{e}_z$, and adopt a continuous medium, real space description of atomic and electromagnetic field operators as described in~\cite{Julsgaard2003}. The electric field of a narrow-band light field of frequency $\omega_L$ is described by Heisenberg operators for orthogonal polarizations $j$ as 
$\hat{\mathbf{E}}=\sum_j(\hat{\mathcal{E}}_j+\hat{\mathcal{E}}_j^{\dagger})$, with
$\hat{\mathcal{E}}_j(z,t)=g\mathbf{e}_j\frac{1}{\sqrt{2\pi}}\int\hat{a}_{k,j}e^{i(kz-\omega_L t)}dk$, where $g=\sqrt{\hbar\omega_L/2\epsilon_0 A}$ scales the field strength per photon, and $\mathbf{e}_j$ are unit polarization vectors. This can be written as
\begin{equation}
   \hat{\mathbf{E}}_j(z,t)=g[\hat{a}_j(z,t)\mathbf{e}_j+\hat{a}^{\dagger}_j(z,t)\mathbf{e}^*_j].
\end{equation}
Here, the creation and annihilation operators are defined as density amplitudes in position space, obeying $[\hat{a}_i,\hat{a}^{\dagger}_j]=\delta_{i,j}\delta_z(z)$ for orthogonal polarizations $i,j$, such that $c \hat{a}^{\dagger}_j\hat{a}_j$ describes photon flux. Different light polarizations are conveniently described by introducing Stokes vector components that measure photon flux differences
\begin{align}
\label{eq:Stokesops}
\renewcommand*{\arraystretch}{1.2}
\left(\begin{array}{c}
\hat{S}_x\\
\hat{S}_y\\
\hat{S}_z
\end{array}\right)
=\frac{c}{2}
\left(\begin{array}{c}
\hat{a}^{\dagger}_{x}\hat{a}_{x}-\hat{a}^{\dagger}_{y}\hat{a}_{y}\\
\hat{a}^{\dagger}_{\neswarrow}\hat{a}_{\neswarrow}-\hat{a}^{\dagger}_{\nwsearrow}\hat{a}_{\nwsearrow}\\
\hat{a}^{\dagger}_{+}\hat{a}_{+}-\hat{a}^{\dagger}_{-}\hat{a}_{-} 
\end{array}\right)
=\frac{c}{2}
\left(\begin{array}{c}
\hat{a}^{\dagger}_{+}\hat{a}_{-}+\hat{a}^{\dagger}_{-}\hat{a}_{+}\\
i\hat{a}^{\dagger}_{-}\hat{a}_{+}-i\hat{a}^{\dagger}_{+}\hat{a}_{-}\\
\hat{a}^{\dagger}_{+}\hat{a}_{+}-\hat{a}^{\dagger}_{-}\hat{a}_{-} 
\end{array}\right)
,
\renewcommand*{\arraystretch}{1.0}
\end{align}
where $\hat{a}_{+,-}=(\hat{a}_{x}\mp i\hat{a}_{y})/\sqrt{2}$, $\hat{a}_{\neswarrow,\nwsearrow}=(\pm\hat{a}_{x}+\hat{a}_{y})/\sqrt{2}$, and $\hat{a}_{x,y}$ describe circular $\sigma^{\pm}$, linear $\pm 45\degree$, and horizontal/vertical polarizations, respectively. The Stokes vector components obey commutation rules of angular momentum, i.e., $[\hat{S}_{x}(t),\hat{S}_{y}(t^\prime)]=i\delta(t-t^\prime)\hat{S}_{z}(t)$ and cyclic permutations~\cite{note1}. In addition, we can measure the total photon flux of a beam of power $P$ as $2\hat{S}_0=c(\hat{a}^{\dagger}_{i}\hat{a}_{i}+\hat{a}^{\dagger}_{j}\hat{a}_{j})=\hat{P}/\hbar\omega_L$ using any orthogonal pair $\mathbf{e}_{i,j}$.

We are particularly interested in describing light-matter interaction in the off-resonant regime where absorption of the fields can be neglected. Here, the interaction reduces to spin and polarization dependent dispersion, governed by the frequency dependent polarizability tensor $\hat{\bm{\alpha}}$ of the medium. The interaction energy can be expressed as a second-order perturbation with state-dependent dipole density $\hat{\mathbf{d}}$, i.e., as a light-shift of the atomic ground states. The effective Hamiltonian can be stated as
\begin{align}
\hat{H}_{\mathrm{eff}}&=-\int(\hat{\mathcal{E}}^{\dagger}\hat{\bm{\alpha}}\hat{\mathcal{E}}) A dz=\sum_n\int \frac{\hat{\mathcal{E}}^{\dagger}\hat{\Pi}_g\hat{\mathbf{d}}_n\hat{\mathbf{d}}_n^{\dagger}\hat{\Pi}_g\hat{\mathcal{E}}}{\hbar\Delta_n} A dz,
\end{align}
which sums contributions from transitions to excited states with resonant frequencies $\omega_n$ and corresponding detunings $\Delta_n=\omega_L-\omega_n$. The projector $\hat{\Pi}_g$ reduces the description of atomic dynamics to the relevant ground state manifold. For alkali atoms in their electronic ground state, the atomic dipole moment depends on the individual spin $\hat{\mathbf{F}}_i$, which we describe by a continuous operator function $\hat{\mathbf{f}}(z)$ for dimensionless spin per atom. The collective spin of $N$ atoms distributed over any finite length $l$ with density $\rho(z)$ is expressed as $\sum_{i=1}^N\hat{\mathbf{F}}_i=\int_l\rho(z)\hat{\mathbf{f}}(z)\hbar Adz$. 

Using this description, the effective interaction Hamiltonian for an atomic (sub)ensemble in one of the electronic ground-state hyperfine manifolds ($L=0, J=\frac{1}{2}$) of certain $F$, can be expressed with irreducible tensor components. Following ~\cite{Julsgaard2003,Kupriyanov2005,Geremia2006,DeEchaniz2008} with some corrections, we use the expression
\begin{align}
    \label{eqn:HeffD1}
    &\hat{H}_{\mathrm{eff}}=\frac{g^2}{c}\int_{0}^{L}\hat{\Pi}_F\left\{2\alpha_{F}^{(0)}\hat{S}_0+2\alpha_{F}^{(1)}\hat{S}_z\hat{f}_z\right.\\
    &\quad\left.+\alpha_{F}^{(2)}\left[2\hat{S}_0(\hat{f}_z^2-\hat{f}^2/3)+\hat{S}_+\hat{f}_-^2+\hat{S}_-\hat{f}_+^2\right]\right\}\hat{\Pi}_F\rho A dz,\nonumber 
\end{align}
where $\hat{f}_{\pm}=\hat{f}_x\pm i\hat{f}_y$ and $\hat{S}_{\pm}=\hat{S}_x\pm i\hat{S}_y$. This approximate Hamiltonian depends on the strengths $\alpha^{(k)}_{F}$ of the scalar, vector and tensor components of the polarizability, $k=0,1,2$, respectively. Considering only off-resonant driving of atoms to excited levels with electronic angular momentum $J'$, each component can have up to three contributions from transitions $F\rightarrow F'=F,F\pm 1$, given by
\begin{align}
\label{eq:aterms}
    \alpha_{F}^{(0)}
    =&\mathcal{A}^{F-1}_{F}+\mathcal{A}^{F}_{F}+\mathcal{A}^{F+1}_{F}\\
   \alpha_{F}^{(1)}
   =&\frac{3}{2}\left[-\frac{\mathcal{A}^{F-1}_{F}}{F}-\frac{\mathcal{A}^{F}_{F}}{F(F+1)}+\frac{\mathcal{A}^{F+1}_{F}}{F+1}\right]\nonumber\\
    \alpha_{F}^{(2)}
   =&\frac{3}{2}\left[\frac{\mathcal{A}^{F-1}_{F}}{F(2 F-1)
   }-\frac{\mathcal{A}^{F}_{F}}{F(F+1)}+\frac{\mathcal{A}^{F+1}_{F}}{(F+1)(2F+3)
    }\right]\nonumber.
\end{align}
The three contributions $\mathcal{A}_{F}^{F'}$ for transitions from $F$ to $F'$ are given by the respective detunings together with reduced dipole moments (which, by isotropy convention, sum up three orthogonal polarizations):
\begin{align}\label{eq:Acontributions} 
    \nonumber \mathcal{A}^{F'}_{F}=&\frac{1}{3}\cdot\frac{|\langle J,F||e\mathbf{r}||J',F'\rangle|^2}{\hbar\Delta_{F,F'}}\\
    =&\frac{\pi\epsilon_0 c^3\Gamma_{J'}}{\Delta_{F,F'}\omega_{J'}^3}(2J'+1)(2F'+1)
    \begin{Bmatrix}
        J &J'&1\\
        F' &F & I\\
    \end{Bmatrix}^2.
\end{align}
Here, we used a Wigner 6-j symbol and introduced decay rate $\Gamma_{J'}$ and frequency $\omega_{J'}$ of spontaneous emission from the excited $J'$ levels~\cite{Steck2015}. 
We assume $A_{F}^{F'}=0$ for non-existing transitions with undefined 6-j symbol, and for the mathematically indeterminate case where $F=0$ in the denominator, the higher order terms are $\alpha_{F}^{(1,2)}=0$.

The effective Hamiltonian leads to Heisenberg equations for the Stokes vector $\hat{\mathbf{S}}$, given by $(\partial_t+c\partial_z)\hat{\mathbf{S}}(z,t)=[\hat{\mathbf{S}}(z,t), \hat{H}_{\mathrm{eff}}]/i\hbar$. Neglecting retardation of light as it propagates across short samples, i.e., ignoring the time derivative, and using $[\hat{a}(z,t),\hat{a}^\dagger(z',t')]=\delta(z-z')$,
results in the following propagation equations for the Stokes parameters: 
\begin{align}
 \frac{\partial\hat{S}_x}{\partial z}=&\frac{2g^2\rho A}{\hbar c}\left[-\alpha_{F}^{(1)}\hat{S}_y \hat{f}_z+\alpha_{F}^{(2)}\hat{S}_z(\hat{f}_{\nearrow}^2-\hat{f}_{\nwarrow}^2)\right],\label{eqn:dSxdz}\\
 \frac{\partial\hat{S}_y}{\partial z}=&\frac{2g^2\rho A}{\hbar c}\left[+\alpha_{F}^{(1)}\hat{S}_x \hat{f}_z-\alpha_{F}^{(2)}\hat{S}_z(\hat{f}_x^2-\hat{f}_y^2)\right],\label{eqn:dSydz}\\
\frac{\partial\hat{S}_z}{\partial z}=&\frac{2g^2\rho A}{\hbar c}\alpha_{F}^{(2)}\left[+\hat{S}_y(\hat{f}_x^2-\hat{f}_y^2)-\hat{S}_x(\hat{f}_{\nearrow}^2-\hat{f}_{\nwarrow}^2)\right],\label{eqn:dSzdz}
\end{align}
where we used $45\degree$ rotated operators $\hat{f}_{\nearrow,\nwarrow}=(\pm\hat{f}_x+\hat{f}_y)/\sqrt{2}$ for better clarity. The first two equations, as well as the two terms in the last expression, are each unitary equivalent under a $45\degree$ rotation about the $z$-axis. Terms containing $\hat{S}_0$  in the Hamiltonian, see Eq.~(\ref{eqn:HeffD1}), cause global phase shifts but do not change polarization.
It can be seen that the set of equations describes rotations of the Stokes vector and that the rank-1 and rank-2 components of the polarizability are linked to circular and linear birefringence, respectively.
Linear polarizations $\hat{S}_{x,y}$ experience Faraday rotation about the $z$-axis, proportional to $\alpha_{F,J'}^{(1)}$ and longitudinal spin components $\hat{f}_z$, i.e., atomic orientation along $z$. Similarly, circular polarization $\hat{S}_z$ couples to $\hat{S}_{x,y}$, proportional to $\alpha_{F}^{(2)}$ and the alignment of transversal spin. 

For the case of small optical phase shifts $(\ll 1~\mathrm{rad})$, the induced rotations of the Stokes vector along the atomic ensemble will be small. If we also neglect backaction of light onto the atomic spin on the time scale of light traversion through the sample, we can approximate the right-hand sides of the propagation equations to be constant. 
As a result, the interaction with the atomic ensemble can be described using symmetric collective operators defined as
\begin{align}
    \hat{X}_{x} &= \int\rho(\hat{f}_x^2-\hat{f}_y^2)\hbar^2 Adz=\sum_i(\hat{F}_{x,i}^2-\hat{F}_{y,i}^2)\\
    \hat{X}_{y} &= \int\rho(\hat{f}_{\nearrow}^2-\hat{f}_{\nwarrow}^2) \hbar^2 Adz=\sum_i(\hat{F}_{\nearrow,i}^2-\hat{F}_{\nwarrow,i}^2)\\
    \hat{T}_z &= \int\rho\hat{f}_z\hbar Adz=\sum_i\hat{F}_{z,i},
\end{align}
with corresponding definitions for individual atomic operators $\hat{F}_{j,i}$~\cite{note2}.
With these definitions and approximations, the Stokes operators describing a probe beam after interaction with the atomic ensemble result from the integration of propagation equations as
\begin{align}
\hat{S}_x'&=\hat{S}_x-
g_{F}^{(1)} \hat{S}_y\hat{T}_z+g_{F}^{(2)} \hat{S}_z\hat{X}_y\label{eqn:dSxdzT}\\
\hat{S}_y'&=\hat{S}_y+g_{F}^{(1)} \hat{S}_x\hat{T}_z-g_{F}^{(2)} \hat{S}_z\hat{X}_x\\
\hat{S}_z'&=\hat{S}_z+
g_{F}^{(2)}\left[\hat{S}_y\hat{X}_x-\hat{S}_x\hat{X}_y\right]\label{eqn:dSzdzT},
\end{align}
with coupling constants $g^{(k)}_{F}=2g^2\alpha^{(k)}_{F}/c\hbar^{k+1}$. The first terms in these equations are the Stokes operators for the input light and are responsible for photon shot noise in any polarimeter detection process.

For simplicity, we only consider strong classical probe light that is linearly polarized along the $45\degree$-axes when it enters the atomic ensemble, i.e., $\hat{S}_{x,z}\approx0$ and $\hat{S}_y\approx S_y$. In this case, Eqs.~(\ref{eqn:dSxdz}) and (\ref{eqn:dSzdz}) and corresponding Eqs.~(\ref{eqn:dSxdzT}) and (\ref{eqn:dSzdzT}) reduce to describing circular and linear birefringence independently. The resulting Faraday rotation and resulting ellipticity reduce to 
\begin{align}
\nonumber\hat{S}_x'&=\hat{S}_x-
g_{F}^{(1)} S_y \hat{T}_z\\
\hat{S}_z'&=\hat{S}_z+
g_{F}^{(2)} S_y\hat{X}_x,
\label{eq:birefringence}
\end{align}
with signal strengths proportional to photon flux $S_y$. 
The link between these measurement operators and operators for the collective pseudo-spin formed from a two-level subspace is discussed in
Appendix~\ref{sec:interactionstrength}.

\section{Quantum mechanical interaction strength}
\label{sec:interactionstrength}
In the following, we discuss our measurement scheme in the context of quantum noise and collective interaction strength between light and atoms, with the caveat that we assume the atomic state to be constant during the detection. Quantum mechanical back-action, dynamical phase evolution, as well as redistribution of population due to spontaneous emission into random directions are not included in our theoretical description. The effects on signal noise resulting from back-action and dynamical phase evolution are essentially caused by alternating measurement of non-commuting operators. In principle, they can be circumvented with stroboscopic measurements~\cite{Vasilakis2011,Vasilakis2015} or combined measurements on oppositely oriented ensembles~\cite{Julsgaard2001}. Redistribution of population generally leads to signal loss, but redistribution within or into the probed manifold will also generate additional signal noise.

It is useful to introduce canonical operators for the involved modes of light and atoms. The collective, two-level pseudo-spin is defined by $\hat{J}_j=\frac{1}{2}\sum_i\hat{\sigma}_j$, which sums individual Pauli operators.
For large atom number $n$ and near-symmetric superpositions of the two states, we can define canonical, atomic operators $\hat{x},\hat{p}=\hat{J}_{y,z}/\braket{\hat{J}_x}^{1/2}=\frac{1}{\sqrt{2n}}\sum_i\hat{\sigma}_{y,z}$.  These quadratures obey $[\hat{x},\hat{p}]\approx i$. The variance $\var{\hat{p}}=\frac{1}{2}$ describes the atomic shot noise of level populations $\hat{n}_{1,2}$, for which we can express $\sqrt{2n}\hat{p}=\sum_i\hat{\sigma}_z=\hat{n}_2-\hat{n}_1$ and thus $\var{(\hat{n}_2-\hat{n}_1)}=n$. Similarly, we define operators for light as 
$\hat{y},\hat{q}=\hat{S}_{z,x}/\braket{\hat{S}_y}^{1/2}$, which obey $[\hat{y}(t),\hat{q}(t^\prime)]\approx i\delta(t-t^\prime)$ and correspond to quadratures of the mode that is orthogonally polarized to the classical input beam. 

Based on the analysis described above, we can formulate measurement operators for the detected mode amplitudes. Separate interaction with hyperfine levels is accounted for using atomic projection operators $\hat{\Pi}_{F}=\sum_m\ket{F,m}\bra{F,m}$ in some basis.
Including electronic noise $s(t)$ in the polarimeter signal, the real observables, i.e., including both sides of the symmetric RF spectrum, are then given by 
\begin{align}
    \hat{m}_F&=\frac{1}{\sqrt{2}}\int u^*\left[\vphantom{\sum_i}(s+g_{\mathrm{el}}\hat{S}_z)\right.\\
    &\qquad\qquad +\left.g_{\mathrm{el}}g^{(2)}_FS_{y} \sum_i \hat{\Pi}_{F,i}\hat{\mathbf{F}}^T_i\mathbf{Q}_{\pm} \hat{\mathbf{F}}_i\hat{\Pi}_{F,i}\right]dt\nonumber+h.c.
\end{align}
We rewrite the sum by defining number-like operators  $\hat{n}^F_{l,m}=\sum_i|F,l\rangle_i\langle F,m|_i$, and introduce the RF cycle integrated, atomic operator
\begin{align}
    \hat{\mathbf{Q}}_{FF}=\frac{\omega}{2\sqrt{2}\pi}\int_0^{2\pi/\omega} e^{2i(\omega t+\varphi)}\hat{\mathbf{F}}^T\mathbf{Q}_
\pm\hat{\mathbf{F}} dt+h.c.
\end{align}
to make the approximation
\begin{align}
    \hat{m}_F&\approx \int \frac{u^*+u}{\sqrt{2}}\left(s+g_{\mathrm{el}}\hat{S}_z\right) dt\\
    &\quad + g_{\mathrm{el}}g^{(2)}_F\int\tilde{u} S_{y} dt \sum_{l,m}\bra{F,l}\hat{\mathbf{Q}}_{FF}\ket{F,m}\hat{n}^F_{l,m}, \nonumber
\end{align}
This approximation makes use of the periodicity $\mathbf{Q}_\pm(t)=\mathbf{Q}_\pm(t+2\pi/\omega)$ and is valid for slowly varying envelopes, assuming $\omega>>\gamma,T^{-1}$, which allows for piecewise integration over RF cycles with approximately constant envelope. 

We can scale the expression for our mode amplitudes to canonical operators
\begin{align}
    \hat{p}^F_{l,m}=\frac{\hat{n}^F_{l,m}}{\sqrt{n}},\qquad
    \hat{y}_u=\frac{1}{\sqrt{2}}\int (u^*+u)\frac{\hat{S}_z}{\sqrt{S_y}}dt.
\end{align}
For simplicity, we assume square laser pulses, i.e., constant photon flux $S_y$ over the support of the mode functions. This allows us to introduce detection gain $g_{\mathrm{det}}$ and interaction strength $\kappa_F$ as
\begin{align}
    g_{\mathrm{det}}&= g_{\mathrm{el}}\sqrt{S_y},\quad
    \kappa_F=\hbar^2g^{(2)}_F\sqrt{nS_{y}}\int \tilde{u}dt.
\end{align}
Using the coupling coefficients
\begin{align}
    c^F_{l,m}= \bra{F,l}\hat{\mathbf{Q}}_{FF}\ket{F,m}/\hbar^2,
\end{align}
the resulting mode amplitude can be expressed as
\begin{align}
    \hat{m}_F&=s_u+g_{\mathrm{det}}\left[\hat{y}_u+\kappa_F\sum_{l,m}c^F_{l,m}\hat{p}^F_{l,m}\right],
\end{align}
where the contribution from electronic noise is given by $s_u=\int (u^*+u)s dt/\sqrt{2}$.

For atomic population in only one sublevel $\ket{F,m}$, the relevant quadrature operator will be $\hat{p}_F=\hat{p}^F_{m,m}$, with corresponding coefficient $c_F=c^F_{m,m}=\xi_F(m)h_2(\theta)/\sqrt{2}$. The  expectation value of the measurement is then given by
\begin{align}
    \braket{\hat{m}_F}&=g_{\mathrm{det}}\left(\braket{\hat{y}_u}+\kappa_F c_F\braket{\hat{p}_F}\right).
\end{align}
Using the atomic operator variance
\begin{align}
    \sigma^2_F=\left[\bra{F,m}\hat{\mathbf{Q}}_{FF}^2\ket{F,m}-\bra{F,m}\hat{\mathbf{Q}}_{FF}\ket{F,m}^2\right]/\hbar^4,
\end{align}
and neglecting technical noise in detection gain or coupling strength, 
the variance of the measured mode amplitude becomes
\begin{align}
    \var{\hat{m}_F}&=S_{UU}^{\mathrm{el}}+g_{\mathrm{det}}^2\left[\vphantom{\kappa_F^{(2)}} \var{\hat{y}_u}\right.\label{Eq:variance}\\
    &\qquad\qquad\qquad \left.+\kappa_F^2\left(\frac{\braket{\hat{p}_F}}{\sqrt{n}}\sigma_F^2+c_F^2\var{\hat{p}_F}\right)\right],\nonumber
\end{align}
where we introduced the power spectral density $S_{UU}^{\mathrm{el}}=\var{s_u}$ of electronic noise in the detected voltage $U(t)$ and made use of the fact that the operator $\hat{\mathbf{Q}}_{FF}$ does not change the hyperfine level.

Atomic quantum noise will become relevant in the regime of strong interaction ($\kappa_F\cdot \orderof(F^2)\gtrapprox 1$). For a coherent input state, the light noise is $\var{\hat{y}_u}=\frac{1}{2}$. 
For a symmetric superposition of one state in each hyperfine manifold, i.e., a coherent spin state, the anti-correlated atomic operators each have expectation values $\braket{\hat{p}_F}=\sqrt{n}/2$ and variances $\var{\hat{p}_F}=\frac{1}{4}$. Considering different coupling strengths and detection gains, appropriate weighting of the two mode amplitudes $\hat{m}_{1,2}$ will lead to some effective coupling strength $\tilde{\kappa}\tilde{c}$ and allow for measurements of $\hat{p}=(\hat{p}_2-\hat{p}_1)/\sqrt{2}$ with variance $\var{\hat{p}}=\frac{1}{2}$.
We have to note, however, that the clock states $\ket{1,0}$ and $\ket{2,0}$ used here are generally not eigenstates of $\hat{\mathbf{Q}}_{FF}$, which leads to additional atomic noise contributions according to Eq.~\ref{Eq:variance}.
In the parallel setting under the resonance condition $\theta=\pi/2$, the atomic operator is $\hat{\mathbf{Q}}_{FF}=(\hat{F}_y^2-\hat{F}_z^2)/\sqrt{2}$, providing a true QND measurement of its eigenstate $\ket{1,0}$ with $c_1=1/\sqrt{2}$ and $\sigma_1^2=0$.
Resonant measurement of $\ket{2,0}$ leads to $c_2=3/\sqrt{2}$ and $\sigma_2^2=3/2$.
For general states $\ket{F,F_z=0}$ of bosonic atoms, the additional noise can be calculated from the variance
\begin{align}
    \braket{(\Delta(\hat{F}_{y}^2-\hat{F}_{z}^2))^2}=\frac{(F-1)F(F+1)(F+2)}{8}\hbar^4.
\end{align}

The resulting noise in the combined measurement will depend on the chosen coupling strengths, detection gains and corresponding signal weighting. In principle, a weak measurement of $\hat{n}_2$ is sufficient to gain information on the total atom number $n$ when combined with a strong measurement of $\hat{n}_1$. Therefore, the optimal measurement strategy and achievable degree of measurement induced spin squeezing depends on the uncertainty of total atom number. This analysis together with consideration of back-action, dynamical phase evolution, spontaneous emission, and breakdown of other approximations made throughout the above derivations is beyond the scope of this paper.

From our measurement data we infer operation in the weak coupling regime for the given optical depth. From the ratio $(\tilde{\kappa}\tilde{c})^2\approx n/(\zeta\cdot1.5\times10^{10})$ of assumed atomic shot noise to (aliasing corrected, $\zeta\approx0.5$) photon shot noise equivalent, neglecting electronic noise as well as detector inefficiency, we estimate an effective interaction strength on the order of $\tilde{\kappa}\tilde{c}\gtrsim0.14$ for the measurement of $\hat{p}$ for an experimentally somewhat uncertain atom number $n\approx1.5\times10^8$, which we can compare to the prediction. For long pulses, the interaction is limited by atomic decay. The maximal effective interaction time resulting from an infinite exponential mode function given in Eq.~\ref{eq:modefunction} using $T=\infty$, is $\int \tilde{u}dt=\sqrt{2/\gamma}$. Still assuming constant atomic signal, we can express an upper bound to the coupling strength as
\begin{align}
    \kappa_F&=\frac{\alpha_{F}^{(2)}}{\alpha_{J'}}\Gamma_{J'}\frac{\lambda^2}{4\pi A}\sqrt{\frac{\lambda}{hc}\frac{n P}{2}}\int\tilde{u}dt\\
    &\leq\sqrt{\chi}\cdot\Gamma_{J'}\frac{\lambda^2}{4\pi}\sqrt{\frac{\lambda}{hc}\frac{n}{A}}.
    \label{eqn:explicitkappa}
\end{align}
This defines the detuning dependent figure of merit $\chi=(\alpha_{F}^{(2)}/\alpha_{J'})^2P/A\gamma$ shown in Fig.~\ref{fig:alpha_pol}~c), which determines the maximal signal-to-noise power ratio at fixed optical depth.

For our Gaussian atomic density distribution with standard deviation $\sigma_0\gtrsim1.2~\mathrm{mm}$ and mode matched probe beams, the effective interaction area is  $A=4\pi\sigma_0^2\gtrsim18~\mathrm{mm}^2$~\cite{note3}. With probe detunings and observed, power dependent decay rates for $F=1$ ($\Delta_{1,1}\approx240~\mathrm{MHz}$, $\gamma\approx900~\mathrm{s^{-1}}$ at $P=540~\mu W$) and $F=2$ ($\Delta_{2,1}\approx400~\mathrm{MHz}$, $\gamma\approx350~\mathrm{s^{-1}}$ at $P=120~\mu W$), we predict maximal coupling strengths of $\kappa_1c_1\lesssim0.21$ and $\kappa_2c_2\lesssim0.35$ for $n=1.5\times10^8$ atoms, using long pulses and complete decay. Here, we use shorter pulses, with only $1~\mathrm{ms}$ for the measurement of $\hat{p}_2$. This minimizes expansion of the falling cloud as well as an error on the subsequent measurement of $\hat{p}_1$ due to the increase of population in $F=1$ from spontaneous emission. The short pulse duration reduces the theoretical coupling strength to $\kappa_2c_2\lesssim0.14$. The estimated effective strength compares well with the predicted values. Further increase of coupling strength and entering the strong coupling regime, especially for QND measurements with minimal atomic loss, requires an increase of optical depth.


\begin{thebibliography}{41}
\bibitem{note3}
The quantum mechanical interaction strength and thus signal-to-noise ratio is highest for mode matched beams and is reduced for other profiles or beam irregularities. However, a non-flattop beam shape will interact with the corresponding spatial mode of the atomic ensemble, reducing the effective number of atoms, which becomes relevant for other than coherent spin states.
\end{thebibliography}
\end{document}